\documentclass[11pt,a4paper]{article}

\usepackage[margin=1in]{geometry}
\usepackage{jinstpub}
\usepackage{caption}
\usepackage{subcaption}
\usepackage{amsmath}
\usepackage{listings}
\usepackage{graphicx}
\usepackage{color}

\definecolor{mygreen}{rgb}{0,0.6,0}
\definecolor{mygray}{rgb}{0.5,0.5,0.5}
\definecolor{mymauve}{rgb}{0.58,0,0.82}

\newcommand\BC{\textsf{BambooCore} }
\newcommand\PX{\textsf{PandaXMC} }
\newcommand\UM{\textsf{UserMC} }
\newcommand\UD{\textsf{UserDetector} }
\newcommand\BCl{\textsf{BambooControl} }
\newcommand\mainf{\textsf{main()} }

\title{BambooMC -- A Geant4-based simulation program for the PandaX experiments}
\author[a,b,1]{Xun Chen
  \note{Corresponding author.}}
\author[c]{Chen Cheng}
\author[d]{Mengting Fu}
\author[a]{Franco Giuliani}
\author[a,b,e,2]{Jianglai Liu
  \note{Spokesperson.}}
\author[f]{Xiaoying Lu}
\author[g,3]{Xiangdong Ji
  \note{Spokesperson.}}
\author[a]{Zhicheng Qian}
\author[d]{Hao Qiao}
\author[h]{Qiuhong Wang}
\author[a]{Jingkai Xia}
\author[a]{Pengwei Xie}
\author[a]{Yukun Yao}
\author[a]{Hongguang Zhang}
\affiliation[a]{INPAC and School of Physics and Astronomy, Shanghai Jiao Tong University, MOE Key Lab for Particle Physics, Astrophysics and Cosmology, Shanghai Key Laboratory for Particle Physics and Cosmology,\\ Shanghai 200240, China}
\affiliation[b]{Shanghai Jiao Tong University Sichuan Research Institute,\\ Chengdu 610213, China}
\affiliation[c]{School of Physics, Sun Yat-Sen University, Guangzhou 510275, China}
\affiliation[d]{School of Physics, Peking University,\\ Beijing 100871, China}
\affiliation[e]{Tsung-Dao Lee Institute,\\ Shanghai 200240, China}
\affiliation[f]{School of Physics and Key Laboratory of Particle Physics and Particle Irradiation (MOE),\\ Shandong University, Jinan 250100, China}
\affiliation[g]{Department of Physics, University of Maryland,\\ College Park, Maryland 20742, USA}
\affiliation[h]{Key Laboratory of Nuclear Physics and Ion-beam Application (MOE), Institute of Modern Physics, Fudan University, Shanghai 200433, China}
\emailAdd{chenxun@sjtu.edu.cn}

\abstract{ The purpose of the PandaX experiments is to search for the
  possible events resulted from dark matter particles, neutrinoless
  double beta decay or other rare processes with xenon
  detectors. Understanding the energy depositions from backgrounds or
  calibration sources in these detectors is very important. The
  program of BambooMC is created to perform the Geant4-based Monte
  Carlo simulation, providing reference information for the
  experiments. We introduce the design and features of BambooMC in
  this report. The running of the program depends on a configuration
  file, which combines different detectors, event generators, physics
  lists and analysis packs together in one simulation. The program can
  be easily extended and applied to other experiments.  }

\begin{document}
\maketitle

\section{Introduction}
\label{sec:intro}

The early stages of the PandaX experiments located in the China
Jinping underground laboratory (CJPL)~\cite{Yu-Cheng:2013iaa}, namely
PandaX-I~\cite{Cao:2014jsa} and PandaX-II~\cite{Tan:2016diz}, are
designed to search for the dark matter in the local halo with
dual-phase xenon time projection chambers (TPCs). The physics goals
are extended to search for the coherent elastic scattering of solar
neutrinos, neutrinoless double beta decay (NLDBD) of $^{136}$Xe and
other rare events in the current stage,
PandaX-4T~\cite{Zhang:2018xdp}. Another detector, called PandaX-III,
equipped with a high pressure TPC with enriched gaseous $^{136}$Xe to
search for the NLDBD, is under construction~\cite{Chen:2016qcd}. One
key problem of the experiments is to understand the background level
caused by material radioactivity in the energy region of interest
(ROI), as well as their spatial distribution. It could only be
answered with the
Geant4-based~\cite{Agostinelli:2002hh,Allison:2006ve} full simulation
together with radioactivity levels given by material
screening~\cite{Wang:2016eud}. Another key problem is to understand
the various distributions resulted from the calibration sources. The
program of BambooMC is created to perform the Monte Carlo (MC)
simulation for different types of detectors in PandaX experiments. It
is designed in a modular architecture, in which the different geometry
descriptions of the detector, physics lists, event generators and data
analysis codes can be combined as required with a configuration file
provided by the user. The program has been used in the different
simulation tasks in the PandaX experiments, such as the background
estimation of the PandaX-I, PandaX-II, PandaX-4T detectors as well as
in the early stage design of the PandaX-III detector. It is also used
in the development of a new method to estimate the neutron background
in dark matter search~\cite{Wang:2019opt}.

In this report, we give a detailed description of the features and
application of BambooMC. In section~\ref{sec:overview}, we give an
overview of the program, including the usage and modules. In
section~\ref{sec:core}, the core features provided by the BambooCore are
described. Then we introduce the specialized features for PandaX and
other low background experiments in section~\ref{sec:pandaxmc}. The
extensibility of BambooMC and the possible usage in other experiments
as a starter toolkit is discussed in section~\ref{sec:extend}.  Selected
application of the BambooMC program in the PandaX experiments are
presented in section~\ref{sec:pandaxmc}.  Finally, a brief summary and
conclusion is given in section~\ref{sec:summary}.

\section{Overview of BambooMC}
\label{sec:overview}
The main purpose of BambooMC is to perform Geant4-based MC simulation
with the help of an input configuration file. It can also be used to
export the detector geometry to a GDML (Geometry Description Markup
Language) file when a special command line option of ``-g'' is
provided. Such feature is convenient for the visualization of the
detector geometry in other applications together with the output data,
such as the ROOT analysis framework~\cite{Brun:1997pa}. The typical
usages of BambooMC are given in Appendix~\ref{sec:usage}. In any
cases, the configuration file must be provided.

\subsection{The configuration file}
\label{sec:config}
The configuration file is specified by the command line option of
``-c''. The file content can be orgnized in the format of XML, JSON or
YAML.\footnote{XML is a type of markup language, see
  \url{https://www.w3.org/XML/}. JSON is a data-interchanage format,
  see \url{https://www.json.org}. YAML is a human-friendly language
  for data serialization, see \url{https://yaml.org}.}

The following information may be presented in the top level
entries in the configuration file:
\begin{itemize}
\item the run number (optional);
\item the geometrical information, which includes sub level entries
  about the possible global geometrical parameters, materials,
  combination of detectors and related parameters;
\item the physics list, with name and possible parameters;
\item the event generator, with name and possible parameters;
\item the analysis package, with name and possible parameters.
\end{itemize}

An exemplary configuration file in the YAML format is presented in
Appendix~\ref{sec:config_file}. The run number is set to be 1234. The
final geometry consists of four ``\textsf{detector}''s and a material
construction. The ``\textsf{detector}''s are organized in a
hierarchical manner using the key of ``\textsf{parent}''. The
definition of the keys will be explained in
section~\ref{sec:geometry}. The physics list of
``\textsf{PandaXPhysics}'' and the generator of
``\textsf{SimpleGPSGenerator}'' are used. The analysis pack
``\textsf{PandaXAnalysis}'' with 7 parameters, is used to control the
output and some internal behaviors of the program. Detailed
description of these parts will be given in following sections.

More examples of the configuration file can be found in the
``config'' directory in the public repository.

\subsection{Modules}
\label{sec:design}
With the configuration file, different detector geometries, physics
lists, event generators and analysis modules can be combined to
perform different simulation tasks with a single BambooMC executable.

\begin{figure}[bt]
  \centering \includegraphics[width=0.8\textwidth]{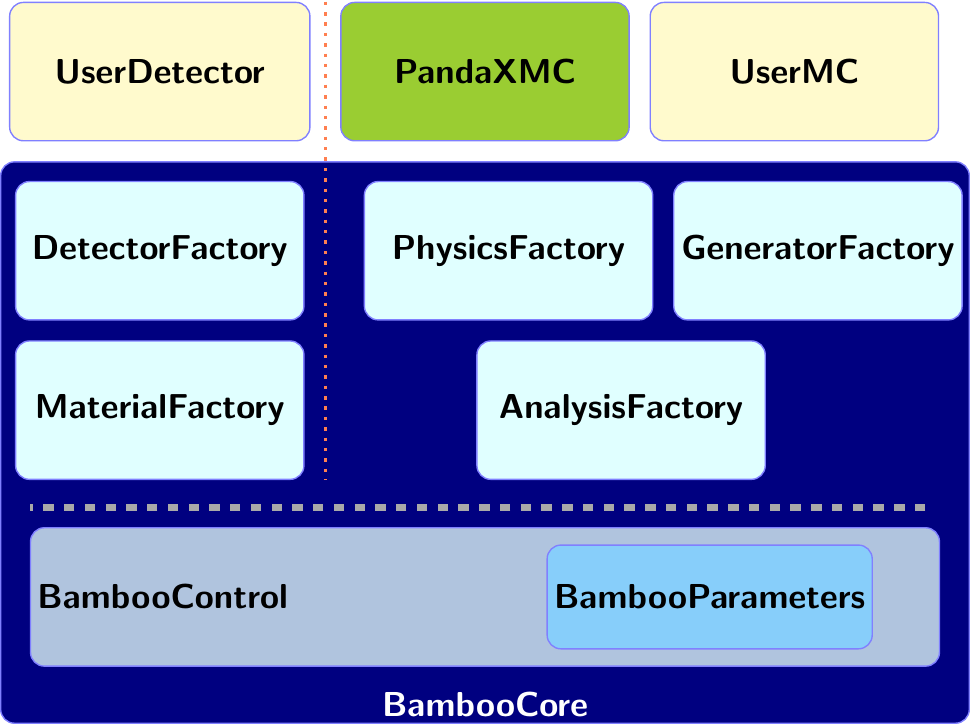}
  \caption{The architecture of the modules in BambooMC. The prefixes of
    ``\textsf{BambooMC}'' are omitted in the ``\textsf{Factory}''
    classes.}
  \label{fig:modules}
\end{figure}

The publicly released program consists of 4 modules, as shown in
Figure~\ref{fig:modules}.
\begin{enumerate}
\item The \BC module provides the \BCl class to parse the
  configuration file and store the different parameters. Different
  types of objects can be created using the so called ``factory''
  design pattern with 5 different ``\textsf{Factory}'' classes.
\item The \PX module defines physics lists, event generators and one
  analysis pack used in the PandaX experiments. The module is built
  on top of the corresponding 3 ``\textsf{Factory}'' classes.
\item The \UM module gives the additional physics lists, event
  generators and analysis packs. It is left for users to extend it
  with new requirements besides the default ones. It has similar
  dependency as that of the \PX module, but the source files are
  placed in a different directory.
\item The \UD module provides detector definition. It is built on top
  of the \textsf{BambooDetectorFactory} and the
  \textsf{BambooMaterialFactory} class.
\end{enumerate}

During the compilation, the \textsf{UserDetector} and \textsf{UserMC}
modules can be customized with the ``-D'' option of CMake, a tool for
build automation.

\subsection{The \mainf  function}
\label{sec:main_function}

The \mainf function defines the workflow of the program. It creates an
object of the \BCl class first, and use it to parse the command line
options and arguments. The supported options and arguments are given
in Appendix~\ref{sec:usage}. The provided configuration file will be
loaded and validated during the process. An object of the
``\textsf{G4UIExecutive}'' class will be created if the ``-i'' option
is given, which means the program will be executed in the interactive
mode. An instance of the ``\textsf{G4RunManager}'' will be created
then. It initializes the detector and physics list through the \BCl
object. The \BCl object will create the generator and analysis pack,
which includes the possible definitions of sensitive detector and
hits, manager of output data, and the objects of
``\textsf{UserAction}''s. The ``\textsf{G4RunManager}'' instance
registers the generator and the created ``\textsf{UserAction}''
objects, and performs initialization thereafter. If the ``-g'' option
is provided together with an output file name, the \mainf function
will dump the geometrical description of the detector to a GDML file
and exit. In other cases, the visualization manager will be created
and initialized. The provided macro file with be parsed by the UI
manager of Geant4. Finally, the program will either start the
simulation with the number of events provided by the command line
option in the batch mode, or start an interactive session in the
interactive mode.

\section{Main features of \BC}
\label{sec:core}
The \BC module provides general functionalities, which can be used in
different experiments.

\subsection{Dynamic geometry construction}
\label{sec:geometry}
The idea of dynamic geometry construction in Geant4 application has
been implemented in many different projects, such as
Mokka\cite{MoradeFreitas:2002kj} and
CLAS12~\cite{Ungaro:2020xlc}. These packages choose to use external
database to store the geometry information and parameters, increasing
the complexity of the deployment and limited their usage.

The \BC module provides a more simple and straightforward way to
define and load the different components of a detector
dynamically. Each component inherited from the \textsf{BambooDetector}
class might provide a ``container'' logical volume as a container for
possible daughter volume. The top-level volumes created in a component
, except for the world volume, should be placed in it parent's
container volume. Different components are loosely coupled and the
configuration file defines the ``real'' detector hierarchy in one
simulation. A simplified example is given in
Listing~\ref{lst:detector}. Each component needs to provide a type,
which is the class name of the component, and a name, which should be
unique among the provided names in the same configuration
file. Non-world component should provide the name of its parent. One
and only one world component is required in a valid configuration
file. A component could appear multiple times in the configuration
file, with different names. In this case, user should be careful with
the parameters provided to these components to avoid possible
overlapping between volumes.

\begin{lstlisting}[frame=single,caption={Example of detector hierarchy in a configuration file of YAML format.}, label={lst:detector}]
  detectors:
  - type: SampleWorld
    name: World
  - type: SampleWater
    name: WaterShield
    parent: World
  - type: SampleSteelContainer
    name: Container
    parent: WaterShield
\end{lstlisting}

The PandaX experiments benefited from this feature greatly. For
example, the PandaX-I and PandaX-II experiments shares the same
passive shielding structure, so the geometrical construction codes for
the shielding can be used in both of the simulations directly. The
codes for the water shielding in CJPL-II are also shared by different
purpose of simulation tasks with different detector constructions,
such as the study of the shielding power and the background
contributed from radon resolved in the water for the PandaX-4T
experiment.

\subsection{Generic parameterization system}
\label{sec:parameter}
BambooMC provides a convenient way for users to customize the program
behavior through its generic parameterization system. Users can
provide parameters in the configuration file for the selected
detectors, generator, physics list and analysis package. Functions to
storage and fetching parameters in the string and arithmetic types,
like \textsf{int} and \textsf{double}, are implemented in the class
\textsf{BambooParameters}. Special parameter values with simple units
provided, like ``\textsf{5.0*m}'', can be parsed with the
``\textsf{evaluateParameter()}'' function.

The most important usage of the generic parameterization system is to
customize the detector construction in combination with the dynamical
geometry construction feature. In the PandaX experiments, such
features are heavily used in the MC simulation on the understanding of
material screening results. Due to the fact that counting samples have
different materials, shape and geometrical parameters, the general
geometry construction codes of the samples are created, leaving the
user to provide the shape, material and related parameters in the
configuration file.

\subsection{Dynamical loading of modular physics list}
\label{sec:physics}
In BambooMC, different modular physics lists can be loaded in
different simulations. A specific physics list is loaded through the
name provided in the configuration file. BambooMC provides an abstract
class of ``\textsf{BambooPhysics}'', which is inherited from
``\textsf{G4VModularPhysicsList}'', to support the dynamical
loading. The name of its concrete daughter class created by users can
be used in the configuration file. The pre-packaged reference modular
physics lists provided by Geant4 can also be used directly by name,
such as ``\textsf{FTFP\_BERT}'' for the simulation of high energy
physics. This feature enables BambooMC to be used easily in
experiments other than PandaX.

\section{The \PX module}
\label{sec:pandaxmc}
Now we introduce the specialized features by the \PX module according
to the requirements of the PandaX experiments.

\subsection{Physics lists for PandaX}
\label{sec:pandax_physics}
The energy ROI of PandaX experiments ranges from sub-keV to a few MeV
level. In the corresponding MC, the background events are mainly from
the decay of radioactive isotopes, resulting in different processes
with the similar range of energies inside the sensitive region of the
detector. The construction of the default physics list for PandaX
experiments, \textsf{PandaXPhysics}, is based on the following
consideration.

First of all, the radioactive decay of isotopes need to be
included. That is been done by registering the
\textsf{G4RadioactiveDecayPhysics} . The decay of other possible
particles, such as muons, is included by registering the
\textsf{G4DecayPhysics}.

For the electromagnetic physics,
the ``Livermore'' implementation (\textsf{G4EmLivermorePhysics}) in
Geant4 is chosen, together with the \textsf{G4EmExtraPhysics} class,
which provides processes related to synchron-radiation of charged
particles.

To simulate the elastic scattering of hadrons, mostly of the neutrons,
\textsf{G4HadronElasticPhysicsHP} is used. The inelastic and capture
processes of neutrons are provided by
\textsf{G4HadronPhysicsShielding}. \textsf{G4StoppingPhysics} is
employed for the possible nuclear capture of negative charged
particles or neural anti-hadrons.

Finally, the class of \textsf{G4IonQMDPhysics} is selected as part of
the physics list, though it may not be used in PandaX simulation.

This physics list is very similar to the pre-packaged modular physics
list of ``\textsf{Shielding}'' provided by Geant4.

Another physics list, \textsf{PandaXOpticalPhysics}, is created on top
of \textsf{PandaXPhysics}, adding the support of
\textsf{G4OpticalPhysics} and enabling the creation of \v{C}erenkov
and scintillation photons. This list is used to study the distribution
of detected scintillation photons for position reconstruction.

\subsection{Event generators}
In the PandaX experiments, the majority of the simulation tasks are
focusing on the background contribution from the radioactive isotopes
in detector materials. These tasks can be done by placing the ions
inside the volume of different part of the detector, and leaving
Geant4 do the rest of works with selected physics lists. We created a
simple wrapper called \textsf{SimpleGPSGenerator} around the Geant4
general particle source (GPS, implemented in the class of
``\textsf{G4GeneralParticleSource}'') to accomplish the function. The
type and spatial distribution of the source ions can be configured
easily by commands supported by GPS. These commands should be placed
in the traditional macro file supported by Geant4, which will be feed
to BambooMC using the command line option of ``-m''.

Additionally, another generator called ``HEPEvtLoader'', which uses
the Geant4 interface to read the HEPEvt format, is created. The
generator can be used in the simulation of collider physics, by
reading the output from generator programs like PYTHIA~\cite{Sjostrand:2014zea}.

\subsection{Output data}
\label{sec:data_output}

The output data by the \PX is organized as the ROOT TTree data format
so that it could be analyzed easily.

By default, the run number and event number are saved in every entry
of the output data. The other variables can be customized in the
configuration file inside the block of ``\textsf{PandaXAnalysis}''.

Two types of sensitive detectors have been defined in \PX. The first
one is used to record the energy deposition of each steps (hits)
inside the sensitive volume. This information is necessary in most of
the simulation tasks to study the background of the PandaX
detector. The related variables in the output data includes:
\begin{itemize}
\item total energy deposited in the volume;
\item number of hits;
\item id number and particle type of each hit;
\item id number and name of the parent particle of each hit;
\item name of the physical process of creation and energy deposition for each hit;
\item position and time of each hit.
\end{itemize}
Another type of sensitive detector is used to record the information
of particle (especially, the gamma) passing through the surface of a
given volume. The output information can be used as input of other
simulations. For example, the stopping effect of the high purity water
shielding system to environment gamma is obtained by the simulations
combining the sensitive detector together with the variant-size water
box. The variables in the output data includes:
\begin{itemize}
\item number of tracks;
\item name and parent name of each track (particle);
\item energy and momentum of each track;
\item position and time of each track.
\end{itemize}
The output of above two types of data can be controlled by the
parameters of ``EnableEnergyDeposition'' and ``EnableSurfaceFlux'' in
the configuration file.

Besides the data generated during the simulation, the information of
the input particles, i.e., the primary particles, can be saved in the
output file by set the value of ``EnablePrimaryParticle'' to be 1. The
related variables include the type, energy, momentum and initial
position of the particle. By default, when no energy deposition or
track is caught by the activated sensitive detectors in an event, the
event will not be saved in the output data. The ``empty'' event can be
forced to be save by set the parameter of ``SaveNullEvents'' to be 1.

The random seed for each events is also saved in the output file. This
variable is useful for debugging. User can repeat the simulation of a
single event by assigning the seed to the parameter of ``UserSeed'' in
the configuration file.

\subsection{Breaking up of long decay chain}
\label{sec:chain_splitting}
Some radioactive isotopes, such as the $^{238}$U and $^{232}$Th, have
a long decay chain and long half-life time at the order of billion
years. To study the background contribution from these isotopes, the
corresponding ion is used as the primary particle in the simulation.
Geant4 will not stop the simulation of an event until the primary ion
and all its descendant particles get into the special ``stop'' states,
such as ``disappeared'', ``out of world'', or being killed explicitly
by user actions. One entry in the output data may contain the
contribution not only from the input isotope, but also from all its
descendants, spanning billions of years. Due to the fact the detectors
record event in a very short time window, the simulation data could
not be used directly. One straightforward solution is to use the
timestamp of each step as the reference of each ``reality''
collisions. But the internal type of time in Geant4 simulation is in
``double'', which is not precise enough for a variable ranging
across more than 15 orders of magnitudes. Another solution is to
``kill'' the radioactive descendant particle in the simulation
explicitly in the user tracking action. This means to study the
contributions from one isotope and its descendants, one need to
perform simulation for each isotope on the decay chain, with greatly
increasing work loads and complexity.

\begin{figure}[bpt]
  \centering
  \includegraphics[width=.47\textwidth]{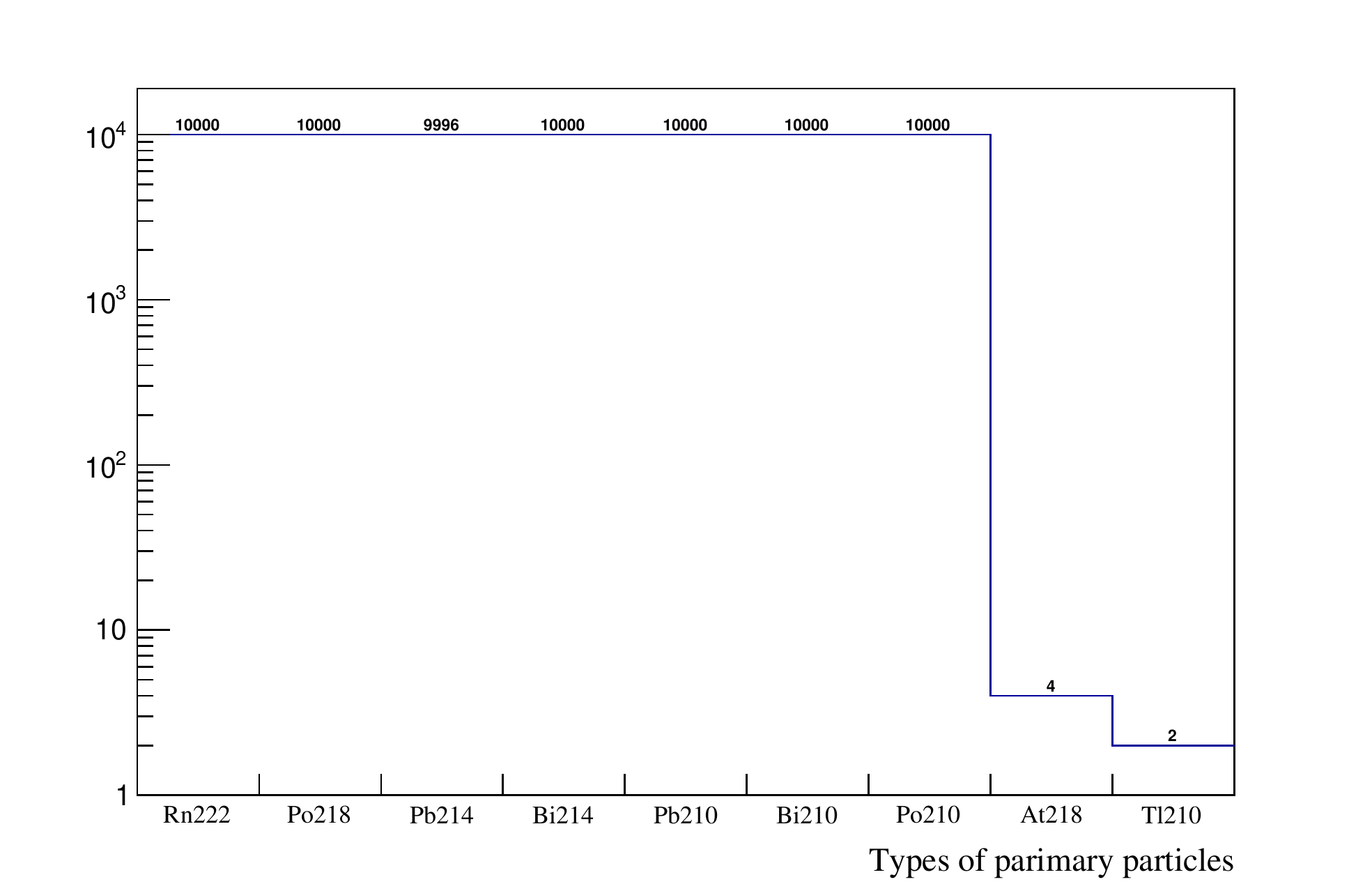}
  \includegraphics[width=.47\textwidth]{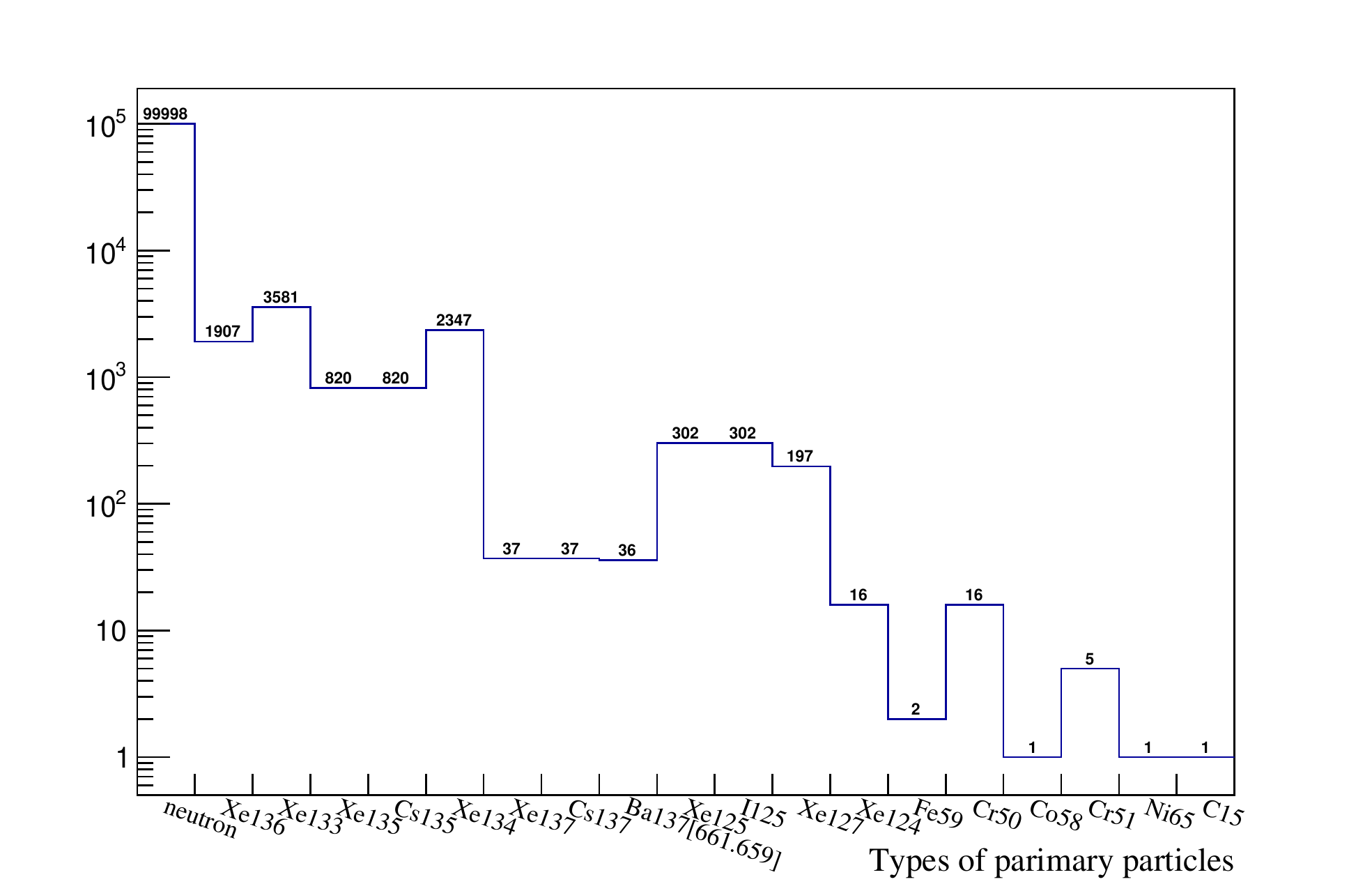}
  \caption{The distributions of primary particles in the simulation by
    enabling the breaking up of long decay chains. Left: Decay
    of $^{222}$Rn resolved uniformly in the xenon target, $10,000$
    events; Right: neutrons of $2.5$ MeV pass through a shell of
    stainless steel and hit the xenon target, $100,000$ events.}
  \label{fig:primary_particle}
\end{figure}

The \PX module provides a mechanism to break up the long decay chain
based on the daughter nuclei and store them in different entries in
the output data. This can be controlled in the configuration file by
setting the value of parameter ``EnableDecayChainSplitting'' to be 1
in the block of ``\textsf{PandaXAnalysis}''.  Once a radioactive ion
is found to be in the ``stop but alive'' state, with zero kinematic
energy, it might be put into the internal particle stack, with time
reset to 0. After tracking of all other particles, the existing hits
information will be extracted and written to the output file as one
entry. Then the tracking of the stacked ion will start again. The
procedures will be repeated until no radioactive particle would be
produced. The hits generated by radioactive isotopes with small half
life, such as $^{214}$Po ($\tau=164.3\mu$s), may be saved together
with those from its parent isotope. This can be customized by setting
a half life threshold for the short-lived isotopes with the parameter
of ``ChainSplittingLifeTime'', which default value is
300$\mu$s. Isotopes with a decay half life smaller than the threshold
will not be separated from their parent. This feature is important to
study the special events like the successive decays of
$^{214}$Bi-$^{214}$Po. This treatment also works for the radioactive
nuclei generated in the process of neutron
capture. Figure~\ref{fig:primary_particle} gives the distributions of
primary particles in the cases of decay of $^{222}$Rn and neutron
scattering, by enabling the breaking up of long decay chains.

\section{Extensibility}
\label{sec:extend}
BambooMC can be extended easily without modifying the structure of the
program, making it useful to be reused in different experiments.

\subsection{Sets and adding classes}
\label{sec:extending}
BambooMC uses the concepts of detector/MC sets to accomplish the
tasks of extending the geometry (physics, generator and analysis), and
the newly created source files will be compiled into the \UD and \UM
modules. A set is a collection of related source files, and will only
be compiled after it is enabled explicitly during the configuration
step with options of ``\textsf{ENABLE\_DETECTOR\_SETS}'' and
``\textsf{ENABLE\_USER\_MC}''. The definition of detectors and
materials should be placed in a detector set, and the definition for
event generator, physics and analysis packs should be placed in a MC
sets.  Multiple detector/MC sets can be enabled simultaneously with
comma separation. Two detector sets (example and optical\_example) and
one MC set (pandax) are shipped with the source code, within the
``user'' directory. To enable these sets, one can configure the
compilation with following command, assuming that ``..'' is the root
of the BambooMC source tree:
\begin{lstlisting}
  cmake -DENABLE_DETECTOR_SETS=example,optial_example \
    -DENABLE_USER_MC=pandax ..
\end{lstlisting}

BambooMC also provides three scripts to help the generation of
extended codes. These functions and the usage of the scripts are given
below.
\begin{itemize}
\item \textsf{add\_detector.pl}: to generate the definition of a sub-class
  of ``\textsf{BambooDetector}'' in a given detector set. The set name and the
  class name should be provided.
\item \textsf{add\_material.pl}: to generate the definition of a
  sub-class of ``\textsf{BambooMaterial}'' in a given detector
  set. The set name and the class name should be provided.
\item \textsf{add\_mc\_class.pl}: to generate the definition of a
  sub-class of either ``\textsf{BambooGenerator}'',
  ``\textsf{BambooPhysics}'' or ``\textsf{BambooAnalysis}'' in a given
  MC set. The set name, type of class and the class name should be
  provided.
\end{itemize}
The scripts will generate the class definitions, with codes related to
the factory method in the source files so that they can be used in the
configuration file. The scripts will also add CMake rules to compile
the created source files.

\subsection{The pandax set in \UM}
\label{sec:um_pandax}
Besides the generators, physics list and analysis package in the \PX
module, following functionalities are provided in the pandax set
within \UM.

An event generator called ``\textsf{EventLoader}'' is provided to read
the modified output of the Decay0 package~\cite{Ponkratenko:2000um},
which contains information of electrons generated in double beta decay
events, and create electrons as primary particles in the
simulation. User can specify the shape of the spatial distribution of
the electrons and confine them in given physical volumes. The
generator is useful when studying the signals from the rare double
beta decay events.

Some simulation tasks in the PandaX experiments require the simulation
of optical photons. The detection of photon and output of the photon
information, including the position, time, energy, and creation
process, is packed in the \textsf{PandaXOpticalAnalysis}.

\subsection{BambooMC as a starter toolkit}
\label{sec:start}

BambooMC is widely used in the different tasks in the PandaX
experiments, with different combination of geometric definitions,
generators, physics lists and analysis packs. Actually, the features
provided by BambooMC makes it to be an ideal starter toolkit for
different particle experiments.

For example, in the early design stage of an experiment, only simple
events with a single type of particle are required. So the
\textsf{SimpleGPSGenerator} can work. The optimization of the final
detector always includes the frequent updating of the size of
different components of the detector. User can create class for each
of the components, use free parameters to control their properties,
and combine them in the configuration file. By updating the
configuration file, different detector concepts can be simulated and
tested. Most of the underground experiments can use the
\textsf{PandaXPhysics} directly as their default physics list. The
collider experiments can select one of the pre-packaged physics lists
provided by Geant4. In most of the experiments, information of energy
deposition is the only required output data from the Geant4 MC
simulation, thus the output from the \textsf{PandaXAnalysis} can be
used. By using of BambooMC as the starter toolkit, a lot of
repeated work can be avoided.

\section{Selected applications in the PandaX experiments}
\label{sec:pandaxmc}
BambooMC is widely used in the PandaX experiments for different
purpose, including the calculation of the detection efficiency of the
counting stations, estimation of the background contribution from the
detector materials in various PandaX detectors, study of event
distributions of different calibration sources, and the correlated
production of high energy gammas and neutrons in detector
materials. Since the background estimation results and the comparison
with calibration data have been described in other
publications~\cite{Chen:2016qcd, Cui:2017nnn, Zhang:2018xdp,
  Wang:2020coa}, we only introduce how the BambooMC program is used in
the estimation of detection efficiency for the counting stations and
the improvement of neutron event generator.

\subsection{Detection efficiency of the counting station}
\label{sec:counting}
Two high purity germanium (HPGe) counting stations for the material
screening of PandaX are operational in CJPL currently. A detailed
introduction to the station equipped with an Ortec GEMMX HPGe detector
(GEMMX) is presented in Ref.~\cite{Wang:2016eud}. Another station
equipped with a Canberra BE3830 HPGe detector (BE3830), was firstly
set up in the Minhang campus of Shanghai Jiao Tong
Univesiy~\cite{Wang:2017phd}, and was delivered to CJPL in 2019. Both
stations have a similar structure, with a passive shield constructed
with lead and copper. The HPGe detector is placed inside a chamber
enclosed by the shield and cooled by liquid nitrogen with a specially
designed cold finger. Samples to be screened are placed on top of the
detector inside the chamber. The HPGe detector measures the energy
spectrum of a sample. By subtracting the background spectrum without
samples, the characteristic gamma energy peaks resulted from the
radioactive isotopes in the sample can be recognized. The event number
within a specified peak can be used to estimate the activity of the
corresponding nuclei by considering the time of screening and the
detection efficiency.

\begin{figure}[bt]
  \centering
  \includegraphics[width=0.6\textwidth]{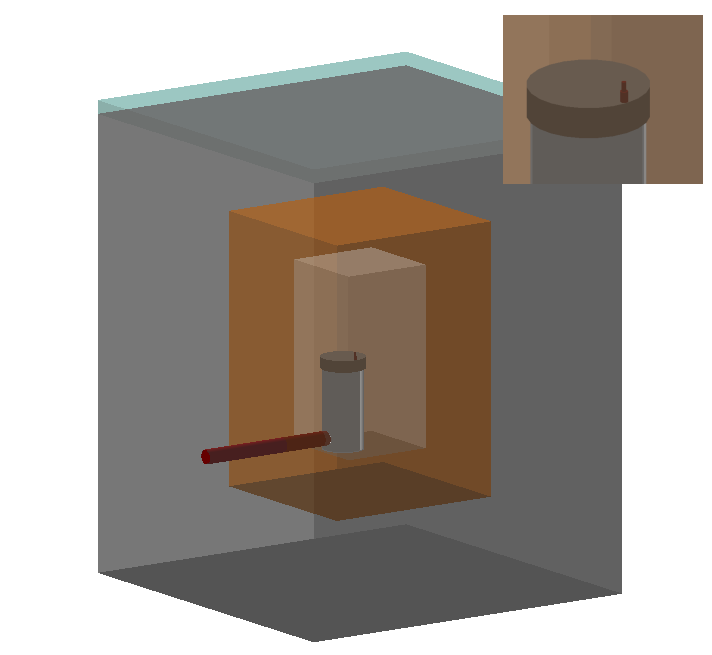}
  \caption{Simulated geometry of the BE3830 station, with a sample on
    the top edge of the cylinder shaped HPGe detector, which is placed
    in the chamber enclosed by the shield of lead (grey) and copper
    (brown). A zoom-in view of the sample is shown in the top right.}
  \label{fig:counting_station}
\end{figure}

Due to the fact that different samples have different shapes and
materials, the detection efficiency can only be estimated with the MC
simulation. The works are carried out with BambooMC. The default
physics list of \textsf{PandaXPhysics}, the generator of
\textsf{SimpleGPSGenerator}, and the analysis package of
\textsf{PandaXAnalysis} are used in the simulation.  A detector set of
``counting'' is created for the geometric description of the stations
and samples.  Two classes of \textsf{CSGEMMXDetector} and
\textsf{CSBEDetector} are created for the two stations GEMMX and
BE3830, respectively. Container logical volumes are provided by the
classes so that the geometric description of different samples can be
placed in. The simulated geometry of the BE3830 station with a sample
is shown in Figure~\ref{fig:counting_station}. For the samples with
regular shapes, several classes are created, with material names and
geometric parameters dynamically loaded from the configuration
files. The dynamic geometry construction feature of BambooMC
simplified the simulation works for the sample counting in PandaX.  A
standard water resolved source of $^{60}$Co and $^{137}$Cs with known
activities is used to validate the simulation
results~\cite{Wang:2016eud,Wang:2017phd}. The relative difference
between the detection efficiencies obtained from simulation and
measurement is smaller than $5\%$, and is accounted as one of the
systematic errors.

\subsection{Improved neutron generator for dark matter search}
\label{sec:neutron}
In the liquid xenon detectors, single scattering nuclear recoil events
generated by neutron can not be distinguished from those generated by
the weakly interactive massive particles (WIMPs). Thus the precision
estimation of neutron backgrounds is critical for these
experiments. The main sources of neutron are from the ($\alpha$,$n$)
reactions and spontaneous fission of radioactive nuclei in the
detector materials. 

In the conventional treatment, the package of a modified
SOURCES-4A~\cite{Wilson:1999so, Tomasello:2008ri}, which can calculate
the rate and energy spectrum of neutrons for given radioactive nuclei
in given type of materials, is served as the neutron generator in
subsequent simulations. But the correlated production of neutrons and
gammas are not considered in this approach, leading to the
over-estimation of neutron background. An improved approach to
estimate the neutron background is developed for the PandaX-II
experiment~\cite{Wang:2019opt,Wang:2020phd}, using the BambooMC
program.

\begin{figure}[hbt]
  \centering
  \begin{subfigure}[t]{0.45\textwidth}
    \includegraphics[width=\textwidth]{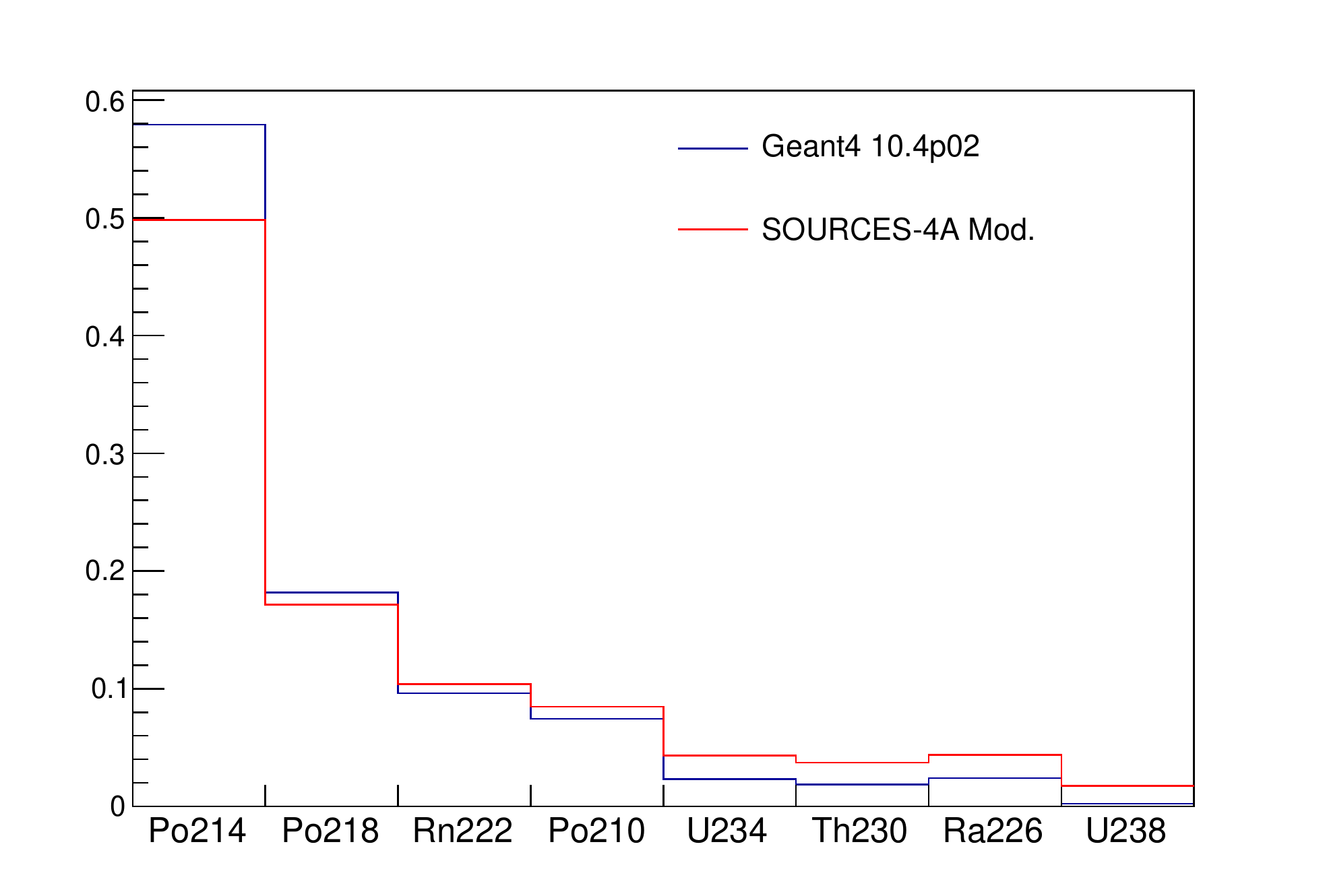}
    \caption{Fraction of neutron contribution nuclei in the ($\alpha$,
      $n$) process.}
  \end{subfigure}
  \begin{subfigure}[t]{0.45\textwidth}
    \includegraphics[width=\textwidth]{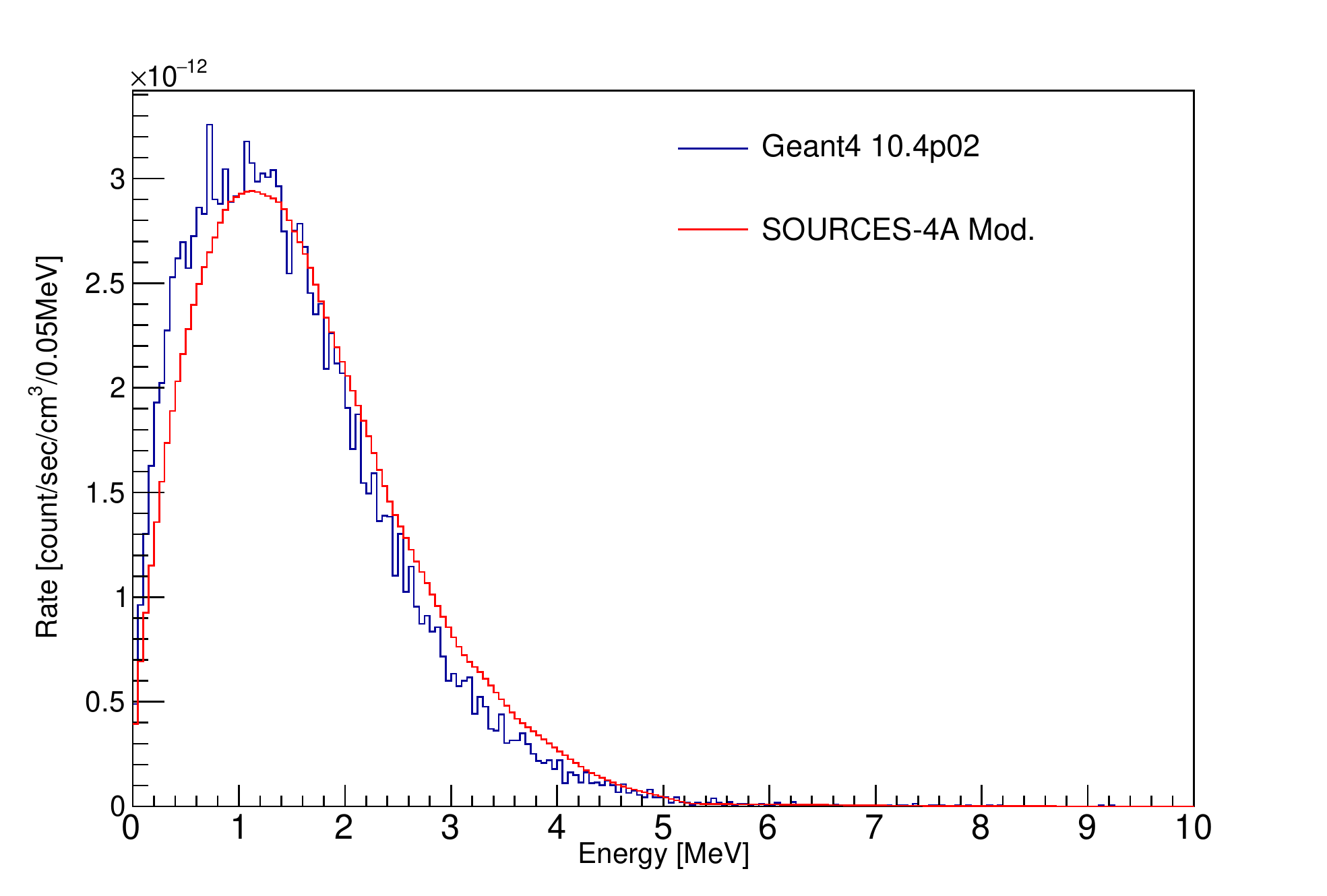}
    \caption{Energy spectrum of neutrons.}
  \end{subfigure}
  \caption{Comparison of neutron production of the $^{238}$U chain in
    PTFE, using the simulation of Geant4 and the SOURCES-4A
    calculation.}
  \label{fig:neutron_comp}
\end{figure}

The new approach employs Geant4 to perform the simulation of the
($\alpha$, $n$) and spontaneous fission processes of radioactive
nuclei in different materials, and uses the output from the simulation
as the input of rest simulations. For example, to study the neutron
produced resulted from $^{238}$U in the polytetrafluoroethylene
(PTFE), a simple geometry of a small PTFE sphere, with a radius of
0.1~mm, is created, and $^{238}$U sources are sampled in the center of
the sphere, using the \textsf{SimpleGPSGenerator}. Then the
information of neutron and gamma tracks passing through the surface of
the sphere is recorded by the \textsf{PandaXAnalysis}. A modified
version of the \textsf{PandaXPhysics}, which uses the Li\`{e}ge
cascade model \textsf{INCL++}~\cite{Mancusi:2014fba} to modeling the
($\alpha$, $n$) process, is used in the BambooMC simulation, with the
ABLA evaporation model enabled~\cite{Heikkinen:2008zz}. By assuming
the activity of $^{238}$U is 1~mBq/kg inside the PTFE and the secular
equilibrium of the $^{238}$U decay chain, the neutron production rate
from the Geant4 simulation (version 10.04p02) is $7.00\times10^{-11}$
s$^{-1}\cdot$cm$^{-3}$, which is at the same order of the calculation
of $1.37\times10^{-10}$ s$^{-1}\cdot$ cm$^{-3}$ from SOURCES-4A.  The
detailed comparison of the fraction of neutron contributors and the
neutron energy spectra are shown in Fig~\ref{fig:neutron_comp}, with
all the spectra normalized to those from SOURCES-4A. The neutron
properties from the two approaches are consistent with each other.

In the PandaX experiments, different simulations with the $^{238}$U,
$^{235}$U, and $^{232}$Th chains in different materials are carried
out using BambooMC. The correlated information of the neutrons and
gammas in the output files are extracted out, and used as primary
particles in the subsequent detector simulations, to provide a more
realistic estimation of the neutron background.

\section{Summary and Conclusion}
\label{sec:summary}

We have developed a modular simulation framework, BambooMC, based on
the widely used Geant4 toolkit, for the PandaX experiments. It has
been used in many simulation tasks to study the background
contribution, signal properties, or other aspects in these
experiments. The generic parameterization system of the program
provides a new way to customize the behavior of the simulation. User
can freely change the simulation without rebuilding the program.  The
design of the program enables it to be extended easily and to be used
in other types of experiments.

The current version of BambooMC is 2.0, which can be found in the
online GitHub repository of
\url{https://github.com/pandax-experiments/BambooMC}. The program is
delivered under GUN Public License version 3. We hope it could benefit
the community of particle experiments by reducing the possible repeat
works.

\section*{Acknowledgment}
\label{sec:ack}
We thank Dr. Xiang Liu for the early stage simulation program for
PandaX-I and the helpful discussion, Mr. Yubo Zhou, Mr. Wenbo Ma for
the building test of the program in different environments. This work
is supported by the grants from National Science Foundationof
China(Nos. 11505112, 12090060), a grant from the Ministry of Science
and Technology of China (No. 2016YFA0400301). We thank the Office of
Science and Technology, Shanghai Municipal Government
(No. 11DZ2260700, No. 16DZ2260200, No. 18JC1410200) and the Key
Laboratory for Particle Physics, Astrophysics and Cosmology, Ministry
of Education, for important support. We also thank the sponsorship
from the Chinese Academy of Sciences Center for Excellence in Particle
Physics (CCEPP).

\bibliographystyle{JHEP}
\bibliography{refs}

\appendix
\section{Usage of BambooMC}
\label{sec:usage}
User need to provide command line options to have the BambooMC to run.
The supported command line options are:
\begin{itemize}
\item ``-c'': need an argument to specify the input configuration file.
\item ``-m'': need an argument to specify the Geant4 macro file.
\item ``-i'': run in interactive, no argument required.
\item ``-n'': need an argument to specify the number of events to be simulated.
\item ``-o'': need an argument to specify the name of the output ROOT file.
\item ``-g'': need an argument to specify the output GDML file.
\end{itemize}

Following are some examples:
\begin{lstlisting}[language=bash,frame=single,caption=The program will simulate 10000 events and write the output to file ``output.root'']
BambooMC -c config.json -m radon.mac -n 10000 -o out.root
\end{lstlisting}

\begin{lstlisting}[language=bash,frame=single,caption=The program will run in interactive mode]
BambooMC -c config.json -i
\end{lstlisting}

\begin{lstlisting}[language=bash,frame=single,caption=The program will dump the detector geometry into an GDML file ``out.gdml'']
BambooMC -c config.json -g out.gdml
\end{lstlisting}

\section{Examples of configuration file}
\label{sec:config_file}
\begin{lstlisting}[frame=single,caption=An example configuration file in YAML format.]
run: 1234
geometry:
  material:
    name: SampleMaterial
  detectors:
  - type: SampleWorld
    name: World
    parameters:
      half_x: 4*m
      half_y: 4*m
      half_z: 4*m
  - type: SampleWater
    name: WaterShield
    parent: World
    parameters:
      width_x: 4*m
      width_y: 4*m
      width_z: 4*m
  - type: SampleSteelContainer
    name: Container
    parent: WaterShield
    parameters:
      radius: 1.05*m
      height: 2.1*m
  - name: XenonDetector
    type: SampleCylinder
    parent: Container
    parameters:
      radius: 1*m
      height: 2*m
physics:
  name: PandaXPhysics
  parameters:
    cutlength: 0.1*mm
generator:
  name: SimpleGPSGenerator
analysis:
  name: PandaXAnalysis
  parameters:
    EnableEnergyDeposition: 1
    EnableSurfaceFlux: 0
    EnablePrimaryParticle: 1
    SaveNullEvents: 0
    EnableDecayChainSplitting: 1
    ChainSplittingLifeTime: 400*us
    UserSeed: 0
  
\end{lstlisting}

\end{document}